\documentclass[journal]{IEEEtran}
\usepackage{glossaries}
\usepackage{amsmath}
\usepackage{amsfonts}
\usepackage{multirow}

\usepackage{times}
\usepackage{floatflt}
\usepackage{graphics}
\usepackage{graphicx}
\usepackage{subfigure}
\newglossary{symbols}{sym}{sbl}
\makeglossary
\newglossaryentry{fn}{type=symbols, name={\ensuremath{ff}},sort=fn,description={this is test}}
\ifCLASSINFOpdf
\else
\fi

\begin{document}

\makeatletter
 \def\hlinewd#1{%
   \noalign{\ifnum0=`}\fi\hrule \@height #1 \futurelet
    \reserved@a\@xhline}
 \makeatother
%
\title{An Estimation Method Using Periodic Inspection of Indicators}
%
%
%

\author{Zheng~Wang


\thanks{Zheng Wang was with the National Institute of Standards and Technology, Gaithersburg,
MD, 20899 USA, e-mail: zhengwang98@gmail.com.}
}

\maketitle

\begin{abstract}
This paper proposes a new approach for estimating the failure time distribution using the indicator data. The indicators, which are checked by periodic inspection of a standby redundant system, only convey whether at least one failure occurs per interval. The estimation procedure first obtains the estimation of the forward recurrence time using the indicator data. Then the mean is estimated based on its relationship with the forward recurrence time. And the estimation of the sampled Cdf is thus derived based on its relationship with the forward recurrence time and the mean. Finally, the Cdf function is estimated using interpolation method. The simulation results showed that the estimation method performed well for the four Weibull distributions.
\end{abstract}


%
\IEEEpeerreviewmaketitle

\section*{Notation}
\begin{tabular}{p{2.5cm}p{5.5cm}}
$X_i$ & inter-event~times, $i=1,2,...$; they are i.i.d. random variables\\
$F(*)$ & Cdf of $X_i$\\
$W$ & time from the origin to the next event (forward recurrence time)\\
$g(*)$,$G(*)$,$\overline{G}(*)$ & pdf, Cdf, Sf of $W$\\
$\mu$ & $\textrm{E}\{X\}$, mean failure time\\
$A_i$ & number of events in the interval $( ( i-1) t , it]$, $i=1,2,...,v$\\
$v$ & number of observation intervals\\
$a_i$ & observed value of $A_i$\\
$\mathbb{A}_i$ & indicator function of $A_i$, $\mathbb{A}_i\equiv1$, if $A_i=0$; $\mathbb{A}_i\equiv0$, otherwise\\
$\mathbb{a}_i$ & observed value of $\mathbb{A}_i$\\
$T$ & observation period $(0, vt]$\\
$t$ & fixed interval for observations\\
\end{tabular}
\section{Introduction}

Periodic inspection is one of the most common testing or maintenance activities, which are basically designed for estimating or improving the reliability of a system \cite{PI_1},\cite{PI_2},\cite{PI_3},\cite{PI_4}. With the purpose of detecting failures, periodic inspection is typically undertaken regularly at a constant rate while the system is continuously operated. That means the system is not interrupted by the failures, thereby the inter-failure times are not impacted by the inspection activities. Compared with continuous inspection, the failures can be detected not immediately but only at the next inspection once they occur. In other words, there is some delay between the real occurrence of a failure and its detection.

The uninterruptibility of the system is ensured at least in two cases:
\begin{itemize}
  \item The system has enough redundancy so that upon failure of one component in function, the backup is automatically activated as a replacement in a negligible amount of time. Such redundant systems were well studied and documented \cite{RS_1},\cite{RS_2},\cite{RS_3}.
  \item The system has components with hidden or soft failures. While so-called hard failures make the system stop functioning as soon as they occur, the system can continue to operate when one or more soft failures take place. Those components with soft failures include protective components, standby components, and secondary components, which do not carry out the main functions of a system \cite{SF_1},\cite{SF_2}.
\end{itemize}

Despite of the uninterruptibility of the system, the consumption of redundant components or the presence of components with soft failures, if accumulated over time and not detected nor addressed, can still eventually reduce the performance or even break the operation of the system. For example, the system may fail if the redundant components are used up, or the number of components with soft failures amount to a certain threshold. So periodic inspection are performed to detect and\/or rectify the failures and to ensure that the system is safe and reliable. Typical rectification efforts taken by periodic inspection includes remedying the loss of redundancy and fixing the components with failures.

A strong assumption on the data provided by periodic inspection would be the count data,  number of failures per interval. Dattero and White \cite{PI} proposed an estimation approach based on periodic inspection of the count data. However, the count data is not always known or provided in periodic inspection. In a wide diverse of systems, there is no failure counter dedicated for the inspection. Some inspections, especially those performed passively or by outsiders, are much like black-box testing where only limited or even minimum information about failures can be retrieved. This paper considers an indicator on which the estimation of the failure-time distribution is based. The indicator only tells whether one or more failures occur per interval rather than the exact number of failures. The data provided by indicator may be most readily available for the monotonically evolved parameters of a system.

For example, since the accumulated failure repair time of a system is non-decreasing, a change in the accumulated failure repair time from the previous inspection would serve as the indicator that there is one or more failures in the interval (but the number of failures are never known because the repair time of an individual failure may vary).The example is illustrated in Fig.1, where the $i+1$ th, $i+2$ th, and $i+4$ th inspections observe an increase in the accumulated failure time caused by one, two, and one failure(s) respectively. By comparison, the $i+3$ th inspection finds no change in the accumulated failure time in Fig.1, indicating no failure occurs between the $i+2$ th inspection and the $i+3$ th inspection.

The incapability of obtaining the number of failures may result from the privacy concerns or the lack of continuous inspections. Some systems ensure the privacy of their operations by leaking a minimum information where some sensitive data such as the number of failures are kept private and unavailable for outsiders. Some systems cannot afford the cost of continuous inspections so that the failures are not one-by-one thoroughly identified and counted.

This paper provides an estimation method using periodic inspection of indicators. Based on the indicator function of failures observed at fixed time intervals, the failure-time distribution is estimated.First,the estimation process is formally developed. Then the estimation procedure is stated. Finally, the performance of the estimation method is assessed using simulated data for various Weibull distributions.

\begin{figure}[!t]
\centering
{\includegraphics[width=0.9\linewidth]{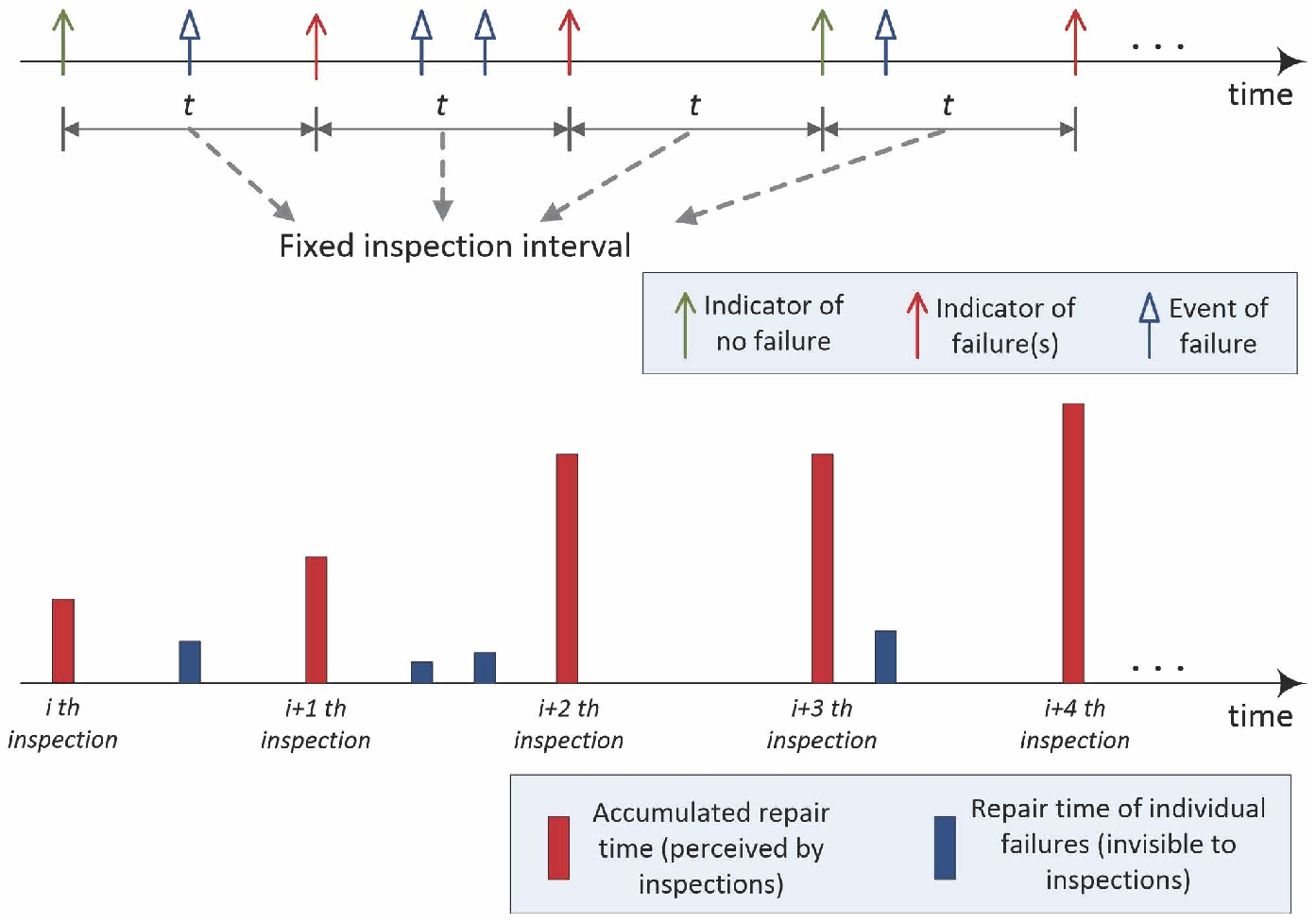}}
\begin{center}
\footnotesize{Fig. 1. Periodic inspection and an example of accumulated failure repair time as the indicator}
\end{center}
\end{figure}


\section{Problem Definition}
Assumptions:
\begin{itemize}
  \item There is a stationary renewal process.
  \item The inter-event times are i.i.d. random variables.
  \item Data are collected on a standby redundant system over a fixed number of collection intervals.
  \item The goal is to find an estimator for the Cdf of inter-event times.
\end{itemize}
The equation for the stationary renewal process to be estimated is \cite{Renew}
\begin{equation}
F(x)=1-\mu g(x)
\end{equation}

\section{Estimation Procedure}

Given (l), a reasonable estimator for $F(x)$ would seem to be:
\begin{equation}\tilde{F}(x)=1-\hat{\mu}\hat{g}(x)
\end{equation}
In developing $\tilde{F}$, difficulties arise, however, because only event occurrence for each interval is known; neither the actual inter-event times $X_l$, $X_2$, ... nor the event counts for each interval are known.

\subsection{Estimator for $\mu$, $g(*)$, $G(*)$, $\overline{G}(*)$ at $kt$}

If the event counts for each interval are known, the mean inter-event time, $\mu$, can be straightforwardly estimated by dividing the total period, $T$, by the sum of the observed event counts. However, independently estimating $\mu$ is more much difficult, since the event counts for each interval are not known. So estimating $\mu$ may not be fully decoupled from estimating $g(x)$. Instead this paper estimates $\mu$ and $g(x)$ simultaneously. The associated Sf is assessed at $kt$ as these are the only points where the data provide useful information. The Sf, $\overline{G}(*)$ is first estimated using

\begin{equation}
\Pi(kt)=\sum_{i=1}^{v-k+1}B_k(i)/(v-k+1)~~~k=1,2,...,K
\end{equation}

Notation:

\begin{tabular}{p{2cm}p{6cm}}
$\Pi(0)=1$ & \\
$B_k(i)$ & an indicator; $B_k(i)\equiv\prod_{j=i}^{i+k-1}\mathbb{A}_j$\\
$\Pi(kt)$ & proportion of time no events are reported over $k$ consecutive intervals each of length $t$\\
$\hat{p}(kt)$ & an observation of $\Pi(kt)$\\
$b_k(i)$ & an observation of $B_k(i)$
\end{tabular}

Note that a) $B_1(i)=1$ if $A_i=0$, b) overlapping intervals
are used (for $k=2,3,...,K$) in (3), and c) $K$ is selected such
that $K\leq v$.

The estimator $\Pi(kt)$ has at least three desirable properties:
\begin{itemize}
  \item It is an unbiased estimator (even though the $B_k(i)$ terms are not $s$-independent, the mean value of the sum is still
equal to the sum of the mean values of each $B_k(i)$ term and E\{$B_k(i)$\}=Pr\{$W>kt$\} for all $i$)..
  \item It has a recursive method of calculation (since $B_k(i)=B_{k-1}(i)B_{k-1}(i+1)$ for $k=2,3,...$).
  \item It is monotonically nonincreasing in $k$.
\end{itemize}

For formal proofs of these properties and some more esoteric results, see \cite{Event}.
The $g(kt)$ is estimated using the centered difference equation to estimate a derivative for $k=1,2,...,K-1$:
\begin{equation}\nonumber \hat{g}(kt)=[\hat{p}((k-1)t)-\hat{p}((k+1)t)]/2t~~~kt>0
\end{equation}
\begin{equation}
\hat{g}(0)=1/\hat{\mu}
\end{equation}
\begin{equation}
\lim_{x\to\infty}\hat{g}(x)=0
\end{equation}
since these properties hold for $g(x)$. To ensure: $\lim_{x\to\infty}\hat{g}(x)=0$,
the $K$ is usually chosen in such a way that $\hat{p}((K-2)t)=\hat{p}((K-1)t)=\hat{p}(Kt)=0$.
Using $\hat{g}(kt)$ from (4), $\tilde{F}(kt)$ from (2) is not necessarily monotonic. Therefore, $F(kt)$ is estimated using
\begin{equation}
\tilde{F}(kt)=\textrm{MAX}\{\tilde{F}(kt),\tilde{F}((k-1)t)\}
\end{equation}
\begin{equation}
\tilde{F}(0)\equiv0
\end{equation}
Using this procedure ensures that the sequence $\tilde{F}(t),\tilde{F}(2t),...,\tilde{F}((K-1)t)$ is monotonically nondecreasing and that $0\leq\tilde{F}(kt)\leq1$, $k=1,2,...,K-1$.

According to (4), we need to derive $\hat{g}(0)$ in order to have $\hat{\mu}$. We have

\begin{equation}
\int_{0}^{\infty}g(x)=1
\end{equation}
which can be written as

\begin{equation}
\sum_{k=1}^{\infty}\int_{(k-1)t}^{kt}g(x)=1
\end{equation}
The integrals in (9) can be estimated using $\hat{g}(kt)$

\begin{equation}
\sum_{k=1}^{\infty}[\hat{g}(kt)+\hat{g}((k-1)t)]t/2=1
\end{equation}
Since $\hat{g}(kt)=0$, for $k=K-1,K,...$, we have

\begin{equation}
\sum_{k=1}^{K-1}(\hat{g}(kt)+\hat{g}((k-1)t))t/2=1
\end{equation}
So $\hat{g}(0)$ can be derived from the following

\begin{equation}
\hat{g}(0)t/2+\sum_{k=1}^{K-2}\hat{g}(kt)t=1
\end{equation}
Then $\tilde{F}(kt)$ can be estimated per (2).

\subsection{Estimator for $g(*)$, $G(*)$, $\overline{G}(*)$ at $x$}

The final step in the procedure is the estimation of this function for $x>0$, but
$x\neq t,...,(K-1)t$. The recommended procedure is linear interpolation so that for $(k-1)<t<kt$:
\begin{equation}
\tilde{F}(x)=\tilde{F}((k-1)t)+(x-(k-1)t)[\tilde{F}(kt)-\tilde{F}((k-1)t)]/t
\end{equation}

Linear interpolation has the advantages of simplicity and ease
of use. However, any other interpolation method can be used,
providing it gives a monotonic estimate of $F(x)$.

\subsection{Algorithm}
For the sake of clarity, we give an algorithm that pulls together the segments discussed above and given in (2) - (12).

\noindent 1. CALCULATE
\begin{itemize}
        \item $v=T/t$
\end{itemize}

\noindent 2. DEFINE
\begin{itemize}
        \item $\hat{p}(0)=1$
        \item $\hat{F}(0)=0$
\end{itemize}

\noindent 3. FOR $k=1,2,...,v$
\begin{itemize}
        \item For $i= 1,2,...,(v-k+1)$\\
        If $k=1$ SET $b_1(i)=1$ IF $a_i=0$ OR SET $b_1(i)=0$ IF $a_i>0$\\
        If $k>1$ CALCULATE $b_k(i)= b_{k-1}(i)b_{k-1}(i+1)$
        \item CALCULATE $\hat{p}(kt)=\sum_{i=1}^{v-k+1}b_k(i)/(v-k+1)$
\end{itemize}

\noindent 4. DETERMINE
\begin{itemize}
        \item $K=\mathop{\arg\min(x)}\limits_{\substack{x=1,2,...,v-1\\\hat{p}((x-2)t)=\hat{p}((x-1)t)=\hat{p}(xt)=0}}$
\end{itemize}

\noindent 5. FOR $k=1,2,...,K-1$
\begin{itemize}
        \item CALCULATE $\hat{g}(kt)=[\hat{p}((k-1)t)-\hat{p}((k+1)t)]/2t$
        \item CALCULATE $\hat{\mu}=t/[2(1-\sum_{k=1}^{K-2}\hat{g}(kt)t)]$
        \item CALCULATE $\tilde{F}(kt)=1-\hat{\mu}\hat{g}(kt)$
        \item DETERMINE $\tilde{F}(kt)=\textrm{MAX}\{\tilde{F}(kt),\tilde{F}((k-1)t)\}$
\end{itemize}

\noindent 6. FOR $x>0$, but $x \neq t, 2t, ...,(K-1)t$
\begin{itemize}
        \item LINEARLY INTERPOLATE FOR $(k-1)t<x<kt$\\
        $\hat{F}(x)=\hat{F}((k-1)t)+(x-(k-1)t)[\hat{F}(kt)-\hat{F}((k-l)t]/t$
\end{itemize}

\section{Performance Evaluation}
While the most desirable evaluation is based on the closed form analytic results of the estimation method, the complexity of the joint distributions involved in the proposed estimation procedure makes it hard to derive the closed form expression. For example, few closed form analytic results are readily available for the joint distributions of the indicators. The computation of $\hat{g}(kt)$ and $\hat{F}(x)$, especially when compound with the indicators, adds to the complexity.

Since the statistical complexity of the proposed estimation procedure results in the hardness of the closed form analytic evaluation, we rely on Monte Carlo simulation for evaluations.
The Weibull distribution has a variety of distribution shapes and wide use in reliability studies. So we choose four Weibull distributions in the simulations, among which one special case is the exponential distribution. As illustrated in Table 1, the four Weibull distributions all have their means around 1.

\begin{table}[!htb]
\centering
\caption{Parameters of Weibull Distribution\protect\\$F(x)=1-exp[-(x/\alpha)^\beta]$}\label{tab:1}
\begin{tabular}{lccc}
\hlinewd{1pt}
& Scale & Shape\\
Distribution & $\alpha$ & $\beta$ & Mean\\
\hlinewd{1pt}
1 & 1.090 & 5.0 & 1.001\\
2 & 1.009 & 3.5 & 0.908\\
3(Exponential) & 1.000 & 1.0 & 1.000\\
4  & 0.878 & 0.8 & 0.995\\
\hlinewd{1pt}
\end{tabular}
\end{table}

For each of the four distributions, we generated 1000 independent runs of
event epochs. For each run, estimates were made for periods $T$ of 50, 100, 500, 1000.
Each period $T$ was sectioned into periodic intervals of fixed length $t$ of 0.1, 0.2, 0.5, 1 and the
indication data $t$ for each interval were recorded. We obtained the Cdf estimation for each run,
for each period of length $T$, and for each interval of length $t$.

The performance metric used was the maximum absolute distance between the estimated Cdf and actual Cdf, $|\hat{F}(x)-F(x)|$. Thus, in cases where the measure is small, the estimation procedure closely approximates the actual Cdf. As the intermediate result of the final estimation, the estimated mean may also serve as a parameter interested and desired in many cases. So we also used the absolute distance between the estimated mean and actual mean, $|\hat{\mu}-\mu|$, as the other performance metric. When this measure is small, the estimation procedure successfully approximates the actual mean.

Table 2 summarizes the means of the maximum absolute Cdf differences for the various combination of run length, $T$, and $t$ for each of the distributions. In general, we can see that the estimation method shows good performance for all combinations of setting. Distribution (1) and Distribution (2) have similar best performance (Distribution (1) is better for some settings and Distribution (2) is better for the other settings). The estimation method performed worst for Distribution (4). The explanation may be that the estimation method is relatively less favorable for distributions that have high probability mass concentrated in a small interval. This comparative performance fall is caused by two operations in the estimation procedure: 1) the derivation of $\hat{g}(0)$ (thus $\mu$) based on $\hat{g}(kt)$ assuming that the integral of the probability distribution of the forward recurrence time can be approximated using $\hat{g}(kt)$; 2) the linear interpolation which is a sub-optimal interpolation method for some distributions.

\begin{table}[!htb]
\centering
\caption{Means of Maximum Absolute Cdf Difference}\label{tab:2}
\begin{tabular}{cccccc}
\hlinewd{1pt}
&  &\multicolumn{4}{c}{Distribution}\\
\cline{3-6}\\
$T$ & $t$ & (1) & (2) & (3) & (4)\\
\hlinewd{1pt}
50 & 0.1 & .078 & .090 & .101 & .123\\
   & 0.2 & .078 & .071 & .119 & .163 \\
   & 0.5 & .181 & .142 & .200 & .269\\
   & 1 & .331 & .315 & .321 & .373\\
\\
100 & 0.1 & .055 & .056 & .081 & .110\\
   & 0.2 & .065 & .052 & .096 & .153 \\
   & 0.5 & .178 & .132 & .197 & .261\\
   & 1 & .325 & .314 & .313 & .371\\
   \\
500 & 0.1 & .028 & .056 & .081 & .110\\
   & 0.2 & .053 & .036 & .091 & .153 \\
   & 0.5 & .168 & .130 & .197 & .265\\
   & 1 & .323 & .309 & .316 & .367\\
   \\
1000 & 0.1 & .022 & .019 & .050 & .095\\
   & 0.2 & .052 & .034 & .091 & .152 \\
   & 0.5 & .168 & .130 & .196 & .265\\
   & 1 & .323 & .309 & .317 & .367\\
\hlinewd{1pt}
\end{tabular}
\\
\begin{tabular}{ccccccc}
&  &  &  &  &  &  \\
\multicolumn{7}{c}{Factor Means}\\
\hlinewd{1pt}
Distribution & Mean & $T$ & Mean & $t$ & Mean & Grand Mean\\
\cline{1-2}
\cline{3-4}
\cline{5-6}
\cline{7-7}
\\
(1) & .152 & 50 & .185 & 0.1 & .072 & .172\\
(2) & .137 & 100 & .172 & 0.2 & .091 &\\
(3) & .173 & 500 & .168 & 0.5 & .192 &\\
(4) & .225 & 1000 & .162 & 1 & .331 &\\
\hlinewd{1pt}
\end{tabular}
\end{table}

In Table 2, we can observe the performance degrade with the increase of the interval $t$ for most of the results. This seems accord with our intuition because more detail is lost for a longer interval. Another finding is that the increase of the observation period $T$ contributes to some improvement of estimation. This also matches our intuitive expectations that more samples facilitates a higher accuracy of estimation. However, a smaller value of $t$ does not always provide a better estimate. For example, Distribution (2) shows a better estimate for $t=0.2$ than for $t=0.1$ when the observation period $T$ is small. The reason of this anomaly mainly lies in the errors introduced by estimating the integral of the probability distribution of the forward recurrence time using $\hat{g}(kt)$ and the sub-optimality of the linear interpolation. The performance of linear interpolation varies for different distributions, so there is some space of improvements if some prior knowledge about the shape of the underlying distribution is known and better interpolation methods can be found accordingly. Another anomaly is no performance improvement with the grow of $T$ in Distribution (3). This phenomena also results from estimating the integral of the probability distribution and the linear interpolation.

Table 3 summarizes the absolute mean differences for the various combination of run length, $T$, and $t$ for each of the distributions. In general, the estimation method performs well for the mean estimation. In terms of the average performance, smaller sampling interval $t$ and longer observation period $T$ both help to enhance the performance. Like the Cdf estimation, some anomalies can be found because of the errors introduced by estimating the integral of the probability distribution of the forward recurrence time using $\hat{g}(kt)$.

\begin{table}[!htb]
\centering
\caption{Means of Absolute Mean Difference}\label{tab:3}
\begin{tabular}{cccccc}
\hlinewd{1pt}
&  &\multicolumn{4}{c}{Distribution}\\
\cline{3-6}\\
$T$ & $t$ & (1) & (2) & (3) & (4)\\
\hlinewd{1pt}
50 & 0.1 & .005 & .013 & .066 & .131\\
   & 0.2 & .007 & .014 & .122 & .188 \\
   & 0.5 & .015 & .033 & .286 & .434\\
   & 1 & .117 & .196 & .597 & .784\\
\\
100 & 0.1 & .004 & .000 & .071 & .109\\
   & 0.2 & .006 & .002 & .094 & .188 \\
   & 0.5 & .010 & .020 & .274 & .395\\
   & 1 & .106 & .195 & .575 & .769\\
   \\
500 & 0.1 & .000 & .003 & .057 & .113\\
   & 0.2 & .001 & .001 & .102 & .183 \\
   & 0.5 & .004 & .018 & .281 & .404\\
   & 1 & .102 & .185 & .583 & .762\\
   \\
1000 & 0.1 & .001 & .000 & .052 & .109\\
   & 0.2 & .001 & .003 & .102 & .188 \\
   & 0.5 & .004 & .017 & .273 & .400\\
   & 1 & .102 & .185 & .581 & .759\\
\hlinewd{1pt}
\end{tabular}
\\
\begin{tabular}{ccccccc}
&  &  &  &  &  &  \\
\multicolumn{7}{c}{Factor Means}\\
\hlinewd{1pt}
Distribution & Mean & $T$ & Mean & $t$ & Mean & Grand Mean\\
\cline{1-2}
\cline{3-4}
\cline{5-6}
\cline{7-7}
\\
(1) & .030 & 50 & .188 & 0.1 & .046 & .178\\
(2) & .055 & 100 & .176 & 0.2 & .075 &\\
(3) & .257 & 500 & .175 & 0.5 & .180 &\\
(4) & .370 & 1000 & .174 & 1 & .412 &\\
\hlinewd{1pt}
\end{tabular}
\end{table}

\section{Conclusion}

This paper proposed a new approach for estimating the failure time distribution using the indicator data obtained by periodic inspections. Simulations showed that the estimation method performed well for the four Weibull distributions. The proposed estimation method can be applied in the system testing or maintenance practices where only the indicator data rather than any more details are available or desired.

\appendices
\end{document}